\documentclass[aps,preprint,superscriptaddress,showpacs,nofootinbib]{revtex4-1}
\usepackage{graphicx}
\usepackage{amsfonts}
\usepackage{mathrsfs}
\usepackage{epsfig}
\usepackage{slashed}
\usepackage{dcolumn}%

\usepackage{ulem}
\usepackage{color}
\usepackage{caption}
\usepackage{subcaption}
\definecolor{My_red}        {cmyk}{0.00,1.00,1.00,0.20}

\usepackage{amsmath}
\usepackage{amssymb}
\usepackage[colorlinks,
linkcolor=blue,
anchorcolor=blue,
citecolor=green
]{hyperref}

\begin{document}

\title{Explanations of the Tentative New Physics Anomalies and Dark Matter
in the Simple Extension of the Standard Model (SESM)}

\author{Tianjun Li}
\email{tli@mail.itp.ac.cn}
\affiliation{
CAS Key Laboratory of Theoretical Physics, Institute of Theoretical Physics,
Chinese Academy of Sciences, Beijing 100190, China
}
\affiliation{
School of Physical Sciences, University of Chinese Academy of Sciences,
No.~19A Yuquan Road, Beijing 100049, China
}
\author{Junle Pei}
\email{peijunle@mail.itp.ac.cn}
\affiliation{
CAS Key Laboratory of Theoretical Physics, Institute of Theoretical Physics,
Chinese Academy of Sciences, Beijing 100190, China
}
\affiliation{
School of Physical Sciences, University of Chinese Academy of Sciences,
No.~19A Yuquan Road, Beijing 100049, China
}
\author{Xiangwei Yin}
\email{yinxiangwei@mail.itp.ac.cn}
\affiliation{
CAS Key Laboratory of Theoretical Physics, Institute of Theoretical Physics,
Chinese Academy of Sciences, Beijing 100190, China
}
\affiliation{
School of Physical Sciences, University of Chinese Academy of Sciences,
No.~19A Yuquan Road, Beijing 100049, China
}
\author{Bin Zhu}
\email{zhubin@mail.nankai.edu.cn}
\affiliation{Department of Physics, Yantai University, Yantai 264005, China}



\begin{abstract}

We revisit the Simple Extension of the Standard Model (SESM), which can account for various tentative new
physics anomalies and dark matter.
The B physics anomalies, muon anomalous magnetic moment, and dark matter can be explained already in SESM. In this work, we study the unitarity constraint on this model and comment the parameter space on the edge of unitarity bounds. Besides, the complete scalar potential is considered which is needed to address the $W$ boson anomaly and more dark matter (co-)annihilation processes are discussed.
We perform a systematic study, and find the viable parameter space which can explain these anomalies
and evade all the current experimental constraints.

\end{abstract}


\preprint{ OSU-HEP-18-nn}

\maketitle

\tableofcontents
\section{Introduction}

After the discovery of Higgs boson at the LHC in 2012~\cite{ATLAS:2012yve,CMS:2012qbp}, the Standard Model (SM) 
has been confirmed to be a correct effective theory at the low energy scale. 
However, we do have a few evidences of new physics Beyond the SM (BSM), for instance,
dark energy, dark matter (DM), neutrino masses and mixings, baryon asymmetry, and inflation, etc~\cite{Davoudiasl:2004be}.
Also, there exist some fine-tuning problems in the SM, for example, gauge hierarchy problem, and strong CP problem, etc.
Thus, the SM is not complete, and we need to explore the new physics.

One promising new physics might happen in flavour sector.
In recent years, the LHCb Collaboration has declared the  anomalies in rare decays of B mesons. The persistent discrepancies between the Standard Model (SM) and the experimental measurements imply that there might be new physics. Such B physics anomalies can be observed in the angular distribution of $B\rightarrow K^{*} \mu^{+} \mu^{-}$ and Lepton Flavor Universality (LFU) ratios $R_{K^{(*)}}=BR(B \rightarrow K^{(*)} \mu^{+} \mu^{-})/BR(B \rightarrow K^{(*)} e^{+} e^{-})$~\cite{Altmannshofer:2021qrr,Aebischer:2019mlg,Albrecht:2018vsa,Li:2018lxi,Bifani:2018zmi,Arnan:2016cpy,Arnan:2019uhr}. 
The recent analyses of the $B^{+} \rightarrow K^{+} \ell^+ \ell^-$ and 
$B^{0} \rightarrow K^{*0} \ell^+ \ell^-$ decays have been performed to  
test the muon-electron universality in two ranges of the square 
of the dilepton invariant mass~\cite{LHCb:2022qnv,LHCb:2022zom}, and the results seem compatible with the SM predictions. However, this is not the final result, since firstly this is just one measurement using a few percent of the data from LHC and with limited precision. Secondly,
although the misidentification backgrounds for electron channel were underestimated in the previous measurement of 
$R_K/R_{K^*}$, the measurements for the muon channels are still correct. 
Thirdly, for the $b\rightarrow s \mu^+ \mu^-$ differential decay branching ratio, there still exist
the theoretical and experimental deviations in the low $q^{2}$ region, and $P^{\prime}_{5}$, etc.
Therefore, we still consider the B physics anomalies in this paper.
For another flavour direction, there exists a 4.2 $\sigma$  discrepancy for the muon anomalous magnetic moment (muon $g-2$) 
$a_\mu = (g_\mu-2)/2$ between the experimental results and theoretical predictions~\cite{Muong-2:2021ojo,Muong-2:2023cdq}.
Although the hadronic contribution might reduce the discrepancy, 
muon $g-2$ is still a promising hint for new physics beyond the SM~\cite{Capdevilla:2020qel,Buttazzo:2020ibd,Capdevilla:2021rwo,Aoyama:2020ynm,Crivellin:2021rbq,Crivellin:2018qmi}, and has been studied extensively~\cite{Li:2021poy,Ahmed:2021htr,Zhu:2021vlz,Calibbi:2021qto,Arcadi:2021cwg}. 
Moreover, new physics might encode in the W mass anomaly. The CDF Collaboration announced a state-of-the-art measurement of the $W$ boson mass, 
which shows 7 $\sigma$ deviation from the prediction of the SM~\cite{CDF:2022hxs}. For recent studies, see Refs.~\cite{Lu:2022bgw,Zhu:2022tpr,Fan:2022dck,Strumia:2022qkt,Athron:2022qpo,Yang:2022gvz,deBlas:2022hdk,Tang:2022pxh,Du:2022pbp,Cacciapaglia:2022xih,Blennow:2022yfm,Sakurai:2022hwh,Yuan:2022cpw,Zhu:2022scj,Fan:2022yly,Liu:2022jdq,Lee:2022nqz,Cheng:2022jyi,Song:2022xts,Bagnaschi:2022whn,Paul:2022dds,Bahl:2022xzi,Asadi:2022xiy,DiLuzio:2022xns,Athron:2022isz,Gu:2022htv,Babu:2022pdn,Heckman:2022the,Balkin:2022glu,Ahn:2022xeq,Han:2022juu,Zheng:2022irz,Nagao:2022oin,Chowdhury:2022moc,Arcadi:2022dmt,Ghorbani:2022vtv,Lee:2022gyf,Benbrik:2022dja,Abouabid:2022lpg,Heo:2022dey,Biekotter:2022abc,Cao:2022mif,Chowdhury:2022dps,Du:2022fqv}. Finally, the DM is  a crucial topic of tentative new physics 
in both particle physics and astronomy, and the observations from astrophysics and cosmology provide 
the overwhelming evidence. The Weakly Interacting Massive Particle (WIMP) provides an excellent DM candidate 
to account for the relic density measured from the observations of Cosmic Microwave Background (CMB), making DM physics a tempting theme.

To address the above anomalies and dark matter,
we revisit the Simple Extension of the Standard Model (SESM)~\cite{Calibbi:2019bay}.
The SESM can already explain the B physics anomaly, muon anomalous magnetic moment, 
 and dark matter simultaneously. In this paper, the following new considerations are involved.  We study the unitarity constraint on this model and present the parameter space on the edge of unitarity bounds. Besides, the complete scalar potential is considered which is needed to address the $W$ boson anomaly and more DM (co-)annihilation processes are discussed.
We perform a systematic study, and find the viable parameter space which can explain these anomalies
and evade all the current experimental constraints.

This paper is organized as follows. In Section~\ref{Field content}, we briefly review the SESM and
present the complete scalar potential. In Section~\ref{SecUnitarity}, the constraints of unitarity on
the parameters in the Yukawa sector are studied. In Section~\ref{Flavor}, 
we explain the above anomalies and dark matter. In Section~\ref{W mass}, the $W$ boson mass anomaly is investigated. 
We conclude in Section~\ref{Conclusion}.

\section{The Simple Extension of the Standard Model}
\label{Field content}
Following Ref.~\cite{Calibbi:2019bay} we introduce a complex singlet scalar $\Phi_{S}$, a  doublet scalar $\Phi_{D}$, 
and two vectorlike pairs of Weyl fermions (that combine into two Dirac fermions) with the same
quantum numbers as the SM quark and lepton doublets $Q^\prime$ and $L^\prime$. Under a discrete $Z_2$ symmetry,
these extra fields are 
all odd while the SM fields are even.
The quantum numbers of exotic fields under SM gauge group are
\begin{equation}
\begin{array}{ccccc}
\hline \text { Field } & \text { spin } & S U(3)_{c} & S U(2)_{L} & U(1)_{Y} \\
\hline Q^{\prime} & 1 / 2 & \mathbf{3} & \mathbf{2} & 1 / 6 \\
L^{\prime} & 1 / 2 & \mathbf{1} & \mathbf{2} & -1 / 2 \\
\Phi_{S} & 0 & \mathbf{1} & \mathbf{1} & 0 \\
\Phi_{D} & 0 & \mathbf{1} & \mathbf{2} & -1 / 2 \\
\hline
\end{array}~.~
\end{equation}

The new vectorlike fermions and scalars can be written as
\begin{equation}
Q^{\prime}=\left(\begin{array}{c}
U^{\prime} \\
D^{\prime}
\end{array}\right)~, \quad L^{\prime}=\left(\begin{array}{c}
L^{\prime 0} \\
L^{\prime-}
\end{array}\right)~, \quad \Phi_{S} \equiv S_{s}^{0}~, \quad \Phi_{D}=\left(\begin{array}{c}
S_{d}^{0} \\
S^{-}
\end{array}\right)~.
\end{equation}

The Lagrangian involving the new fields is given by
\begin{align}
\mathcal{L} \supset &\left(\lambda_{i}^{Q} \overline{Q^{\prime}} Q_{i} \Phi_{S}+\lambda_{i}^{U} \overline{Q^{\prime}} U_{i} \Phi_{D}+\lambda_{i}^{D} \overline{Q^{\prime}} D_{i} \widetilde{\Phi}_{D}+\lambda_{i}^{L} \overline{L^{\prime}} L_{i} \Phi_{S}+\lambda_{i}^{E} \overline{L^{\prime}} E_{i} \widetilde{\Phi}_{D}+a_{H} H^{\dagger} \widetilde{\Phi}_{D} \Phi_{S}\right. \nonumber\\
&\left.+a_{H}^{\prime} H^{\dagger} \widetilde{\Phi}_{D} \Phi_{S}^{\dagger} +  \text { h.c. }\right) -M_{Q} \overline{Q^{\prime}} Q^{\prime}-M_{L} \overline{L^{\prime}} L^{\prime}-M_{S}^{2} \Phi_{S}^{*} \Phi_{S}-M_{D}^{2} \Phi_{D}^{*} \Phi_{D} \nonumber\\
&+ \frac{\lambda_{S}}{2}\left(\Phi_{S}^{+} \Phi_{S}\right)^{2}+\frac{\lambda_{D}}{2}\left(\Phi_{D}^{+} \Phi_{D}\right)^{2} + \lambda_{SD}\left|\Phi_{S}\right|^{2}\left|\Phi_{D}\right|^{2}+\lambda_{S H}\left|\Phi_{S}\right|^{2}|H|^{2}+\lambda_{DH}\left|\Phi_{D}\right|^{2}|H|^{2}\nonumber\\
&+ \lambda_{1}^{\prime} \left(H^{\dagger} \widetilde{\Phi}_{D}\right)\left(\widetilde{\Phi}_{D}^{\dagger} H\right) + \frac{\lambda_{2}^{\prime}}{2}\left(  \widetilde{\Phi}_{D}^{\dagger} H \right)^{2} ~, \label{lagrangian}
\end{align}
where we have systematically considered the scalar potential relative to the phenomenology that we are interested in. We denote the left-handed exotic quarks, right-handed exotic quarks, left-handed exotic leptons, right-handed exotic leptons, left-handed quark doublets, right-handed up-type quarks, right-handed down-type quarks, left-handed lepton doublets, and right-handed down-type leptons as $Q_{L}^{\prime}$, $Q_{R}^{\prime}$, $L_{L}^{\prime}$, $L_{R}^{\prime}$, $Q_{i}$, $U_{i}$, $D_{i}$, $L_{i}$, and $E_{i}$ ($i$=1,2,3), respectively.

After the breaking of the electroweak symmetry, the terms involved $\lambda_{SH}$ and $\lambda_{DH}$ will contribute to masses of $\Phi_{S}$ and $\Phi_{D}$, respectively, while the term involved $a_{H}$ will induce mixing of $S_{s}^{0}$ and $S_{d}^{0}$. For simplicity, we choose $a_{H}^{\prime}$, $\lambda_{1}^{\prime}$ and $\lambda_{2}^{\prime}$ to be zero. 
Therefore, the modified mass matrix of the BSM neutral scalars can be written as
\begin{equation}
U^{\dagger}\left(\begin{array}{cc}
M_{S}^{2}+\frac{\lambda_{SH} v^{2}}{2} & a_{H}^{*} v / \sqrt{2} \\
a_{H} v / \sqrt{2} & M_{D}^{2}+\frac{\lambda_{DH} v^{2}}{2}
\end{array}\right) U=\left(\begin{array}{ll}
M_{S_{1}}^{2} & \\
& M_{S_{2}}^{2}
\end{array}\right)~,
\end{equation}
where $v \simeq 246$ GeV. The corresponding mass eigenstates $S_{1}$ and $S_{2}$ have physical masses of
\begin{align}
    M_{S_{1,2}}^{2}=\left(M_{S}^{2}+\frac{\lambda_{SH} v^{2}}{2}+M_{D}^{2}+\frac{\lambda_{DH} v^{2}}{2} \mp \Delta M^{2}\right) / 2
\end{align}
with
\begin{align}
\quad \Delta M^{2} \equiv \sqrt{\left(M_{D}^{2}+\frac{\lambda_{DH} v^{2}}{2}-M_{S}^{2}-\frac{\lambda_{SH} v^{2}}{2}\right)^{2}+2 a_{H}^{2} v^{2}}~.
\end{align}

The mixing of $S_1$ and $S_2$ is given by
\begin{equation}
U=\left(\begin{array}{cc}
\frac{\sqrt{2} a_{H} v}{\sqrt{\left(M_{D}^{2}+\frac{\lambda_{DH} v^{2}}{2}-M_{S}^{2}-\frac{\lambda_{SH} v^{2}}{2}-\Delta M^{2}\right)^{2}+2 a_{H}^{2} v^{2}}} & -\frac{M_{D}^{2}+\frac{\lambda_{DH} v^{2}}{2}-M_{S}^{2}-\frac{\lambda_{SH} v^{2}}{2}-\Delta M^{2}}{\sqrt{\left(M_{D}^{2}+\frac{\lambda_{DH} v^{2}}{2}-M_{S}^{2}-\frac{\lambda_{SH} v^{2}}{2}-\Delta M^{2}\right)^{2}+2 a_{H}^{2} v^{2}}} \\
\frac{M_{D}^{2}+\frac{\lambda_{DH} v^{2}}{2}-M_{S}^{2}-\frac{\lambda_{SH} v^{2}}{2}-\Delta M^{2}}{\sqrt{\left(M_{D}^{2}+\frac{\lambda_{DH} v^{2}}{2}-M_{S}^{2}-\frac{\lambda_{SH} v^{2}}{2}-\Delta M^{2}\right)^{2}+2 a_{H}^{2} v^{2}}} & \frac{\sqrt{2} a_{H} v}{\sqrt{\left(M_{D}^{2}+\frac{\lambda_{DH} v^{2}}{2}-M_{S}^{2}-\frac{\lambda_{SH} v^{2}}{2}-\Delta M^{2}\right)^{2}+2 a_{H}^{2} v^{2}}}
\end{array}\right)~.
\end{equation}

The couplings between up and down type quarks have a relative misalignment, and we choose
\begin{equation}
\lambda_{i}^{Q_{u}}=\lambda_{i}^{Q}, \quad \lambda_{i}^{Q_{d}}=\sum_{k} \lambda_{k}^{Q} V_{k i}^{*}~,
\end{equation}
where $V$ is the CKM matrix.

In our model, we impose that the lightest BSM particle is $S_{1}$, which is the DM candidate. But the mass hierarchies between $M_{Q^{\prime}}$, $M_{L^{\prime}}$, and $M_{S_{2}}$ are not mandatory.

\section{The Unitarity Constraint}\label{SecUnitarity}

The partial wave amplitude of any 2$\rightarrow$2 scattering process in the high-energy massless limit is~\cite{Allwicher:2021rtd},
\begin{equation}
    a_{f i}^{J}=\frac{1}{32 \pi} \int_{-1}^{1} \mathrm{~d} \cos \theta d_{\mu_{i} \mu_{f}}^{J}(\theta) \mathcal{T}_{f i}(\sqrt{s}, \cos \theta)~,
    \label{1}
\end{equation}
where $\theta$ is the scattering angle in the center of mass frame, $d_{\mu_{i} \mu_{f}}(\theta)$ is Wigner $d$-function, and $\mu_{i/f}=\lambda_{i_1/f_{1}}-\lambda_{i_2/f_{2}}$ with $\lambda_{i_{1/2}}$ and $\lambda_{f_{1/2}}$ standing for the helicities of initial and final particles, respectively.

The unitarity of $S$ matrix respects $S S^{\dagger}=1$. By considering elastic scattering and imposing that the intermediate states are two-particle states, the condition required by the unitarity of $S$ matrix can be obtained.
For general unitarity bound, one has
\begin{equation}
    \operatorname{Re}^{2}\left[a_{i i}^{J}\right]+\left(\operatorname{Im}\left[a_{i i}^{J}\right]-\frac{1}{2}\right)^{2} \leq \frac{1}{4}~.
\end{equation}

At  tree level, the above unitarity bound becomes
\begin{equation}
    \left|\operatorname{Re}\left(a_{i i}^{J, \text { tree }}\right)\right| \leq \frac{1}{2}~.
    \label{3}
\end{equation}

To compute the unitarity bounds in the Yukawa sector, the helicity amplitude approach demonstrated in \cite{Allwicher:2021rtd} is employed, and the general form is given in TABLE \ref{form}. $\mathcal{T}$ is the non-trivial part of $S$ matrix. The absence of $\mathcal{T}$ in $C^{(\prime)}$, $E^{(\prime)}$, and $F^{(\prime)}$ is due to the angular momentum conversation. In the massless limit, the entities in $A^{(\prime)}$ and $B^{(\prime)}$ vanish exactly since $\mathcal{T}$ is proportional to the mass. Besides, the entity in D is present when considering the scattering of two scalars.
\begin{table}[htb]
    \begin{tabular}{c|c|c|c|c} 
    & $\mu_{i}=0$ & $\mu_{i}=0$ & $\mu_{i}=+1$ & $\mu_{i}=+\frac{1}{2}$\quad$\mu_{i}=-\frac{1}{2}$ \\
    &  $++$  $--$ & 00 & $+-$ & $+0$ $-0$ \\
    \hline $\mu_{f}=0$\quad $++$ & $\mathcal{T}^{++++}$ \quad $\mathcal{T}^{++--}$ & $A$ & $B$ & $C$ \\
    \qquad\qquad$--$ & $\mathcal{T}^{--++}$ \quad$\mathcal{T}^{----}$ & & &  \\
    \hline $\mu_{f}=0$\quad $00$ &$A^{\prime}$ &$D$ & $\mathcal{T}^{00+-}$ & $E$ \\
    \hline $\mu_{f}=+1$\quad$+-$ &$B^{\prime}$ & $\mathcal{T}^{+-00}$ & $\mathcal{T}^{+-+-}$ & $F$ \\
    \hline $\mu_{f}=+\frac{1}{2}$\quad$+0$ & & & & $\mathcal{T}^{+0+0}$\qquad \qquad \qquad \\
    $\mu_{f}=-\frac{1}{2}$\quad$-0$ &$C^{\prime}$ & $E^{\prime}$&$F^{\prime}$ & \qquad \qquad \qquad $\mathcal{T}^{-0-0}$
\end{tabular}
\caption{Scattering matrix $\mathcal{T}$.\label{form}}
\end{table}

One can decompose the $\mathcal{T}$ matrix into the following structure
\begin{equation}
    \mathcal{T}_{f_{1} f_{2} i_{1} i_{2}}^{\lambda_{f_{1}} \lambda_{f_{2}} \lambda_{i_{1}} \lambda_{i_{2}}}(\sqrt{s}, \theta)=\bigoplus_{\mathbf{r}} \sum_{m=s, t, u} \mathcal{T}_{m}^{\lambda_{f_{1}} \lambda_{f_{2}} \lambda_{i_{1}} \lambda_{i_{2}}}(\sqrt{s}, \theta) \mathcal{F}_{f_{1} f_{2} i_{1} i_{2}}^{m, \mathbf{r}}(N) \mathbf{1}_{d_{\mathbf{r}}}~,
    \label{4}
\end{equation}
where $\mathcal{T}_{m}^{\lambda_{f_{1}} \lambda_{f_{2}} \lambda_{i_{1}} \lambda_{i_{2}}}(\sqrt{s}, \theta)$ is the Lorentz part and $\mathcal{F}_{f_{1} f_{2} i_{1} i_{2}}^{m, \mathbf{r}}(N)$ is the group factor. To obtain the partial wave amplitude, we should consider a concrete elastic scattering process which can occur in different channels and representations. Our strategy is 
\begin{itemize}
    \item [1)] Considering which partial wave we are interested in;
    \item [2)] Calculating the group factors of the corresponding elastic scattering process;
    \item [3)] Combining Eqs. \ref{1}, \ref{3}, and \ref{4} to obtain the constraint of unitarity on this elastic scattering process.
\end{itemize}
\subsection{The unitarity constraint of $\lambda_{2}^{Q}$}

According to the classification in Ref.~\cite{Allwicher:2021rtd}, the term $\lambda_{i}^{Q} \overline{Q^{\prime}} Q_{i} \Phi_{S}$ in Eq.~\ref{lagrangian} corresponds to the first model of dirac type theory with respect to $SU(3)_C$ and $SU(2)_L$. 
Since $Q_{i}$ denotes a left-handed field, $\overline{Q}^{\prime}$ should be a right-handed field. The term for the second generation is $\lambda_{2}^{Q} \overline{Q}^{\prime}_{R} Q_{2} \Phi_{S}$. In the massless limit, (anti-)fermions described by the left-handed and right-handed spinors have helicities of $-\frac{1}{2}$ ($\frac{1}{2}$) and $\frac{1}{2}$ (-$\frac{1}{2}$), respectively. 
\subsubsection{$J=0$ partial wave}
TABLE.~\ref{form} shows that only $\mathcal{T^{++++}}$ and its conjugation contribute to the $J=0$ partial wave. 
Since $Q_{R}^{\prime}$ and $\overline{Q}_{2}$ are in the fundamental and anti-fundamental representations of $SU(2)_L$ and $SU(3)_C$, the tensor decomposition of $Q_{R}^{\prime} \overline{Q}_{2}$ gives 
\begin{equation}
++\left\{\begin{array}{l}
Q_{R}^{\prime} \overline{Q}_{2} \sim \left( \mathbf{1} \oplus \mathbf{Adj}\right) _{SU(2)_L}\otimes \left( \mathbf{1} \oplus \mathbf{Adj}\right) _{SU(3)_C}~. \\
\end{array}\right.
\end{equation}

The elastic scattering process is mediated by the singlet complex scalar $\Phi_{S}$. Thus, only the channel of the $\mathbf{1}$ representation of $SU(2)_L\times SU(3)_C$ is present.
As for the group factor of this process, we have
\begin{align}
++++\left\{\mathcal{F}_{Q_{R}^{\prime} \overline{Q}_{2} Q_{R}^{\prime} \overline{Q}_{2}}^{s,1}\right.=----\left\{\mathcal{F}_{\overline{Q}_{R}^{\prime} {Q}_{2} \overline{Q}_{R}^{\prime} {Q}_{2}}^{s,1}=N_2\times N_3=2\times 3\right.~,
\end{align}
where $N$ comes from that both fermions involved are in the (anti-)fundamental representation of the $SU(N)$ group (there are more details in Appendix \ref{A}) which means that $N_2=2$ and $N_3=3$ in our case. In the basis of $({{Q}^{\prime}_{R}} \overline{Q}_{2},\overline{Q}^{\prime}_{R} Q_{2})$, based on Eqs.~\ref{1} and \ref{4}, we get
\begin{align}
    a_{S U(3)=1, S U(2)=1}^{J=0}&=\frac{{\lambda_{2}^{Q}}^{2}}{32 \pi} \int_{-1}^{+1} \mathrm{~d} \cos \theta d_{00}^{0}(\theta)\left(\begin{array}{cc}
N_{2} N_{3} \mathcal{T}_{s}^{++++} & 0 \\
0 & N_{2} N_{3} \mathcal{T}_{s}^{----}
\end{array}\right) \nonumber\\
&=\frac{{\lambda_{2}^{Q}}^{2}}{32 \pi} \int_{-1}^{+1} \mathrm{~d} \cos \theta \left(\begin{array}{cc}
2 \times 3 \times(-1) & 0 \nonumber\\
0 & 2\times 3 \times (-1)
\end{array}\right)\\
&=-\frac{{3 \lambda_{2}^{Q}}^{2}}{8 \pi}~,
\end{align}
where the last step means that we find the largest eigenvalue after the diagonalization of the matrix. Besides, we have used $d_{00}^{0}(\theta)=1$ and $\mathcal{T}_{s}^{++++}=\mathcal{T}_{s}^{----}=-1$ according to Table 2 in \cite{Allwicher:2021rtd}. Then, the perturbation unitarity condition of Eq.~\ref{3} leads to the bound
\begin{equation}
    |\lambda_{2}^{Q}| \leq 2.05~.
\end{equation}

\subsection{A brief summary of the unitarity constraints in Yukawa sector}

The elaborate constraints on the parameters of Yukawa sector in our model are given in TABLE \ref{Unitarity summary},
 and the details can be found in Appendix \ref{A}.

\begin{table}[ht]
    \centering
    \begin{tabular}{cccccc}
    \hline
              &$({{Q}^{\prime}_{R}} \overline{Q}_{2},\overline{Q}^{\prime}_{R} Q_{2})$&$({{Q}^{\prime}_{R}} \overline{Q}_{3},\overline{Q}^{\prime}_{R} Q_{3})$&$(L^{\prime}_{R} \overline{L}_{2},\overline{L}^{\prime}_{R} L_{2})$&$(\overline{L}^{\prime}_{L} E, L^{\prime}_{L} \overline{E})$&\\
    \hline
         $J=0$& $|\lambda_{2}^{Q}| \leq 2.05$&$|\lambda_{3}^{Q}| \leq 2.05$&$|\lambda_{2}^{L}| \leq 3.54$& $|\lambda_{2}^{E}| \leq 5.01$ \\
    \hline
            Base I&$(\overline{Q}_{2} \Phi^{\star}_{S},Q_{2} \overline{\Phi}^{\star}_{s})$  &$(\overline{Q}_{3} \Phi^{\star}_{S},Q_{3} \overline{\Phi}^{\star}_{s})$ &$(L^{\prime}_{R} \Phi^{\star}_{S},\overline{L}^{\prime}_{R} \Phi_{S})$ 
              & $(E \widetilde{\Phi}_{D},\overline{E} \widetilde{\Phi}^{\star}_{D})$ \\
    \hline
     $J=\frac{1}{2}$&$|\lambda_{2}^{Q}| \leq 5.01$&$|\lambda_{3}^{Q}| \leq 5.01$&$|\lambda_{2}^{L}| \leq 5.01$&$|\lambda_{2}^{E}| \leq 7.09$\\
    \hline
         Base II&$(Q_{R}^{\prime} \Phi_{s}^{\star},\overline{Q}_{R}^{\prime} \Phi_{s})$  &$(Q_{R}^{\prime} \Phi_{s}^{\star},\overline{Q}_{R}^{\prime} \Phi_{s})$ &$(\overline{L}_{2} \Phi_{S}^{\star},L_{2} \Phi_{S})$ 
               & $(\overline{L}^{\prime}_{L} \widetilde{\Phi}_{D},L^{\prime}_{L} \widetilde{\Phi}_{D}^{\star})$ \\
    \hline
     $J=\frac{1}{2}$&$|\lambda_{2}^{Q}| \leq 5.01$&$|\lambda_{3}^{Q}| \leq 5.01$&$|\lambda_{2}^{L}| \leq 5.01$&$|\lambda_{2}^{E}| \leq 5.01$\\
     \hline
    \end{tabular}
    \caption{Unitarity constraint in Yukawa sector.}
    \label{Unitarity summary}
\end{table}

For $\lambda_{2}^{Q}$, $\lambda_{3}^{Q}$ and $\lambda_{2}^{L}$, the strict constraints come from $J=0$ partial wave, which are $|\lambda_{2}^{Q}| \leq 2.05$, $|\lambda_{3}^{Q}| \leq 2.05$ and $|\lambda_{2}^{L}| \leq 3.54$. While the $J=\frac{1}{2}$ partial wave leads to a relative weaker bound. For $\lambda_{2}^{E}$, the $J=0$ partial wave gives the strict bound $|\lambda_{2}^{E}| \leq 5.01$ as well, however, $J=\frac{1}{2}$ partial wave in an alternative base $(\overline{L}^{\prime}_{L} \widetilde{\Phi}_{D},L^{\prime}_{L} \widetilde{\Phi}_{D}^{\star})$ leads to the same bound.

The unitarity constraints on Yukawa couplings eventually lead the following bounds
\begin{equation}
  \begin{split}
          |\lambda_{2}^{Q}| \leq 2.05 ~,\\
          |\lambda_{3}^{Q}| \leq 2.05 ~,\\
          |\lambda_{2}^{L}| \leq 3.54 ~,\\
          |\lambda_{2}^{E}| \leq 5.01 ~.
  \end{split}
  \label{16}
\end{equation}

Of course, we need to consider the perturbative bounds on the Yukawa couplings, {\it i.e.},
all the Yukawa couplings are smaller than $\sqrt{4\pi}$.

\section{The Tentative New Physics Anoamlies and Dark Matter}\label{Flavor}

In this Section, we will address the $R_{K}$ and $B_{s}$ mixing, muon $g-2$, and dark matter relic density simultaneously.
Random scan of the model parameters is employed, and the benchmark points satisfying all the current constraints are selected.
\subsection{$R_{K}$ and $B_{s}$ mixing}
According to \cite{Calibbi:2019bay}, the contribution to $R_{K}$ can be described by the following effective Lagrangian,
\begin{equation}
\mathcal{H}_{\mathrm{eff}}^{b s \mu \mu} \supset-\mathcal{N}\left[C_{9}^{b s \mu \mu}\left(\bar{s} \gamma_{\mu} P_{L} b\right)\left(\bar{\mu} \gamma^{\mu} \mu\right)+C_{10}^{b s \mu \mu}\left(\bar{s} \gamma_{\mu} P_{L} b\right)\left(\bar{\mu} \gamma^{\mu} \gamma_{5} \mu\right)+\text { h.c. }\right]~,
\end{equation}
where the normalization is given by
\begin{equation}
\mathcal{N} \equiv \frac{4 G_{F}}{\sqrt{2}} \frac{e^{2}}{16 \pi^{2}} V_{t b} V_{t s}^{*}~.
\end{equation}

Contributions to $C_{9,10}^{bs\mu\mu}$ from the BSM particles are
\begin{align}
&\Delta C_{9}^{b s \mu \mu}=-\frac{\lambda_{3}^{Q_{d}} \lambda_{2}^{Q_{d} *}}{128 \pi^{2} \mathcal{N}} \sum_{\alpha=1,2} \frac{\left|U_{1 \alpha}\right|^{4}\left|\lambda_{2}^{L}\right|^{2}+\left|U_{1 \alpha}\right|^{2}\left|U_{2 \alpha}\right|^{2}\left|\lambda_{2}^{E}\right|^{2}}{M_{S_{\alpha}}^{2}} F_{2}\left(\frac{M_{Q}^{2}}{M_{S_{\alpha}}^{2}}, \frac{M_{L}^{2}}{M_{S_{\alpha}}^{2}}\right)~,\label{19}\\
&\Delta C_{10}^{b s \mu \mu}=\frac{\lambda_{3}^{Q_{d}} \lambda_{2}^{Q_{d}} *}{128 \pi^{2} \mathcal{N}} \sum_{\alpha=1,2} \frac{\left|U_{1 \alpha}\right|^{4}\left|\lambda_{2}^{L}\right|^{2}-\left|U_{1 \alpha}\right|^{2}\left|U_{2 \alpha}\right|^{2}\left|\lambda_{2}^{E}\right|^{2}}{M_{S_{\alpha}}^{2}} F_{2}\left(\frac{M_{Q}^{2}}{M_{S_{\alpha}}^{2}}, \frac{M_{L}^{2}}{M_{S_{\alpha}}^{2}}\right)~,
\label{20}
\end{align}
with loop function
\begin{equation}
F_{2}(x, y) \equiv \frac{1}{(x-1)(y-1)}+\frac{x^{2} \log x}{(x-1)^{2}(x-y)}+\frac{y^{2} \log y}{(y-1)^{2}(y-x)}~.
\end{equation}

According to the up-to-date fitting \cite{Altmannshofer:2021qrr} (2$\sigma$), the contribution to LFU violation in $b \rightarrow s \mu \mu$ from $\Delta C_{9}^{b s \mu \mu}$ and $\Delta C_{10}^{b s \mu \mu}$ needs to satisfy
\begin{equation}
    6.74 + 9.04 (\Delta C_{9}^{b s \mu \mu})^2 + \Delta C_{9}^{b s \mu \mu} (14.96 - 10.68 \Delta C_{10}^{b s \mu \mu}) + \Delta C_{10}^{b s \mu \mu} (-13.22 + 11.90 \Delta C_{10}^{b s \mu \mu}) \leq 1~.
\end{equation}

The contribution to $B_{s}-\bar{B}_{s}$ oscillation can be induced from the following operators
\begin{equation}
\mathcal{H}_{\mathrm{eff}}^{b d_{i}} \supset C_{1}^{b d_{i}}\left(\overline{d_{i}} \gamma_{\mu} P_{L} b\right)\left(\overline{d_{i}} \gamma^{\mu} P_{L} b\right)+\text { h.c. } \qquad d_{i}=d,s~.
\end{equation}
The $Q^{\prime}-\Phi_{S}$ box diagram gives Wilson coefficients
\begin{equation}
\Delta C_{1}^{b d_{i}}=\frac{\left(\lambda_{3}^{Q_{d}} \lambda_{i}^{Q_{d}{ }^{*}}\right)^{2}}{128 \pi^{2}} \sum_{\alpha=1,2} \frac{\left|U_{1 \alpha}\right|^{4}}{M_{S_{\alpha}}^{2}} F\left(\frac{M_{Q}^{2}}{M_{S_{\alpha}}^{2}}\right)~,
\end{equation}
where the loop function is
\begin{equation}
F(x) \equiv \frac{x^{2}-1-2 x \log x}{(x-1)^{3}}~.
\end{equation}
The bound given in \cite{DiLuzio:2019jyq,Silvestrini:2018dos} is
\begin{equation}
\Delta C_{1}^{b s}<2.1 \times 10^{-5} \mathrm{TeV}^{-2}~.
\end{equation}

\subsection{Muon g-2}
The contribution of BSM particles in this model to $a_{\mu}$ is
\begin{equation}
\Delta a_{\mu} \approx-\frac{m_{\mu} M_{L}}{8 \pi^{2}} \sum_{\alpha=1,2} \frac{\operatorname{Re}\left(\lambda_{2}^{L} \lambda_{2}^{E *} U_{1 \alpha} U_{2 \alpha}^{*}\right)}{M_{S \alpha}^{2}} f_{L R}\left(\frac{M_{L}^{2}}{M_{S_{\alpha}}^{2}}\right)~,
\end{equation}
where the loop function is
\begin{equation}
f_{L R}(x) \equiv \frac{3-4 x+x^{2}+2 \log x}{2(x-1)^{3}}~.
\end{equation}
The latest $(1\sigma)$ discrepancy is given by \cite{Muong-2:2023cdq}
\begin{equation}
a_{\mu}^{\mathrm{EXP}}-a_{\mu}^{\mathrm{SM}}=(2.49 \pm 0.48) \times 10^{-9}~.
\end{equation}
\subsection{Dark matter phenomenology}

We consider the lightest neutral scalar $S_{1}$ as the DM candidate,
 and the typical freeze-out mechanism controls the DM production. 
In addition, the annihilation and co-annihilation processes can deplete the DM relic density. In our case, we do not consider the neutral vector-like lepton $L_{0}^{\prime}$ to be a DM candidate. Just as mentioned in Ref~\cite{Calibbi:2019bay}, first, the observed relic density requires $M_{L^{\prime}}\approx1.1$ TeV, thus the spectrum would be too heavy to account for flavour observables. Second, since $L_{0}^{\prime}$ interacts with Z boson, the large DM-nucleon scattering cross section cannot be avoided.

\begin{figure}[ht]
	\centering
	\includegraphics[width=0.6\textwidth]{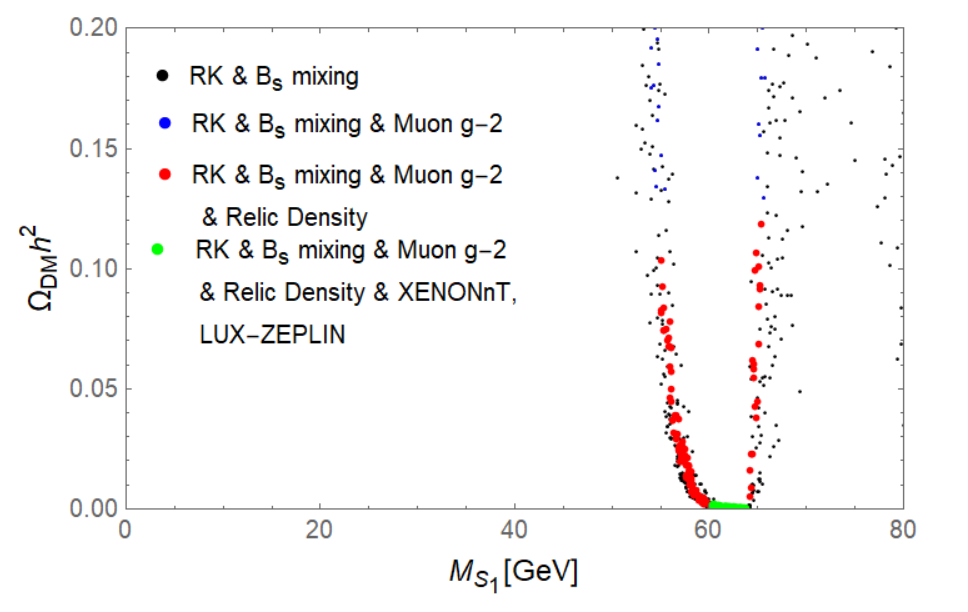}
	\hspace{3em} 
	\captionsetup{justification=raggedright}
	\caption{DM annihilation mediated by Higgs resonance. The black points satisfy the constraints of $R_{K}$ (2$\sigma$) and $B_{s}-\bar{B}_{s}$, the blue points satisfy the constraints of $R_{K}$ (2$\sigma$), $B_{s}-\bar{B}_{s}$ and muon g-2, the red points satisfy the constraints of $R_{K}$ (2$\sigma$), $B_{s}-\bar{B}_{s}$, muon g-2 and DM relic density, and the green points satisfy the constraints of $R_{K}$ (2$\sigma$), $B_{s}-\bar{B}_{s}$, muon g-2, DM relic density, XENONnT~\cite{XENON:2023cxc}, and LUX-ZEPLIN~\cite{LZ:2022lsv} constraints.} 
	\label{higgspole}
\end{figure}

In our random scan, the ranges of parameters have to satisfy the bounds of unitarity given in TABLE \ref{Unitarity summary}. In addition, we set the couplings to the right-handed quarks $(\lambda_{i}^{U}$ and $\lambda_{i}^{D})$ and the left-handed quarks of the first generation $(\lambda_{1}^{Q})$ to be zero, and only consider the second family couplings in the lepton sector ($\lambda_{2}^{L}$ and $\lambda_{2}^{E}$). The elaborate scan ranges of parameters are given in TABLE \ref{RandonScan}.

\begin{table}[ht]
    \centering
    \begin{tabular}{cccccccc}
         \hline 
          $\text { Parameters }$ & $\lambda_{2}^{Q}$& $\lambda_{3}^{Q}$ & $\lambda_{2}^{L}$&$\lambda_{2}^{E}$& $a_{H}/\text{TeV}$ &$\lambda_{S}/2$&$\lambda_{D}/2$ \\
          \hline 
          $\text {Range} $& [-2,2] &[-2,2]&[-3,3]&[-3,3] &[-1,1]&[-3,3]&[-3,3]\\
          \hline 
           $\text { Parameters }$ & $M_{S}$/$\text{TeV}$&$M_{D}/\text{TeV}$&$M_{L^{\prime}}/\text{TeV}$&$M_{Q^{\prime}}/\text{TeV}$&$\lambda_{SH}$ &$\lambda_{DH}$&$\lambda_{SD}$ \\
           \hline 
             \text {Range} & [0,2] &[0,5]&[0.11,5]&[1.2,6]&[-1,1]&[-1,1]&[-3,3] \\
           \hline 
    \end{tabular}
    \captionsetup{justification=raggedright}
    \caption{The scan ranges of parameters employed in the random scan. New parameters of this model 
which are not shown in this table are set to be zero.}
    \label{RandonScan}
\end{table}

FIG.~\ref{higgspole} shows the DM relic density versus the mass of $S_{1}$. The conditions
from $R_{K}$ (2$\sigma$) and $B_{s}-\bar{B}_{s}$ (black dot), the muon g-2 (1$\sigma$) constraint (blue dot), the constraint of (rescaled) DM relic density (red dot), and the constraint from XENONnT and LUX-ZEPLIN experiment (green dot) are applied in order. It can be seen from 
FIG.~\ref{higgspole} that the Higgs mediated DM annihilation channel efficiently depletes the DM, 
leading to the minimum value of relic density.
\begin{figure}
    \centering
    \includegraphics[width=0.9\textwidth]{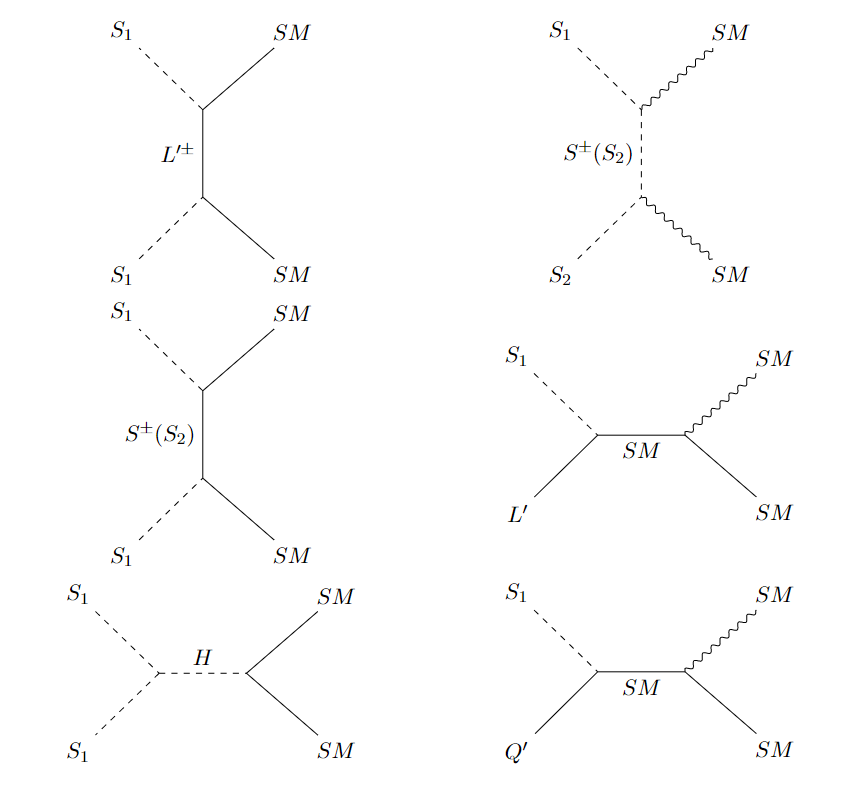}
    \captionsetup{justification=raggedright}
    \caption{The DM (co-)annihilation processes.}
    \label{annihilation}
\end{figure}

The DM relic density is controlled by the (co-)annihilation processes shown in FIG.~\ref{annihilation}.
\begin{itemize}
    \item [1)]  Annihilation of two $S_{1}$ will always happen no matter what the mass differences between $S_{1}$ and other particles are, which are demonstrated in the first column of FIG.~\ref{annihilation}. The annihilation process mediated by the Higgs boson shown in the last diagram of the first column of FIG. \ref{annihilation}  can efficiently deplete the DM relic density.
    \item [2)] Co-annihilations of $S_{1}$ and other particles ($S_{2}$, $L^{\prime}$, and $Q^{\prime}$) will occur when their mass difference are small, which are shown in the second column of FIG.~\ref{annihilation}.
\end{itemize}

Ten benchmark points are given in TABLE \ref{Benchmark points}, and all these points can evade the experimental constraints from $R_{K}$ and $B_{s}$ mixing, and muon g-2. Points 1 and 2 correspond to annihilation of two $S_{1}$.  The Higgs mediated DM annihilation is demonstrated by Points 3 and 4. Point 5 (6), Point 7 (8), and Point 9 (10) are characterized by the co-annihilation between $S_{1}$ and $S_{2}$, $L^{\prime}$, and $Q^{\prime}$, respectively. In addition, the dark matter candidate $S_{1}$ can explain some of the observed relic density and evade XENONnT, LUX-ZEPLIN experiment as well.
\begin{table}[ht]
    \scriptsize
    \centering
    \begin{tabular}{ccccccccccc}
    \hline
         &Point 1 &Point 2 &Point 3 &Point 4 &Point 5 &Point 6 &Point 7 &Point 8 &Point 9 &Point 10 \\
    \hline
           $\text{Relic density}$&0.011522&0.093819&0.000076&0.000046&0.011026&0.01447&0.017495&0.016008&0.04798&0.048167 \\
          $\lambda_{2}^{Q}$&0.10101&0.5996&-0.12509&-0.13069&1.1424&1.3606&0.54168&0.55338&0.22722&0.26312 \\
          $\lambda_{3}^{Q}$&-1.1951&-1.0286&1.8403&1.7261&-1.0692&-0.48519&-1.3764&-1.2747&-1.226&-1.2927\\
          $\lambda_{2}^{L}$&-2.6267&-2.9421&1.9734&1.9598&2.9536&2.9733&-2.492&-2.4106&-2.7338&-2.7466\\
          $\lambda_{2}^{E}$&0.58956&-1.1191&0.18295&0.17041&1.2603&1.3269&-1.9362&-1.9258&-1.5627&-1.5474\\
          $a_{H}\text{/GeV}$&-269.33&105.49&241.84&239.07&77.867&193.76&46.396&48.413&250.22&199.99\\
          $\lambda_{S}/2$&-2.93125&0.88935&0.05591&0.065355&-2.36415&1.4342&-1.2625&-1.3191&-2.0845&-2.08535\\
          $\lambda_{D}/2$&-2.8127&1.41265&0.25705&0.276025&1.3104&-2.75535&-1.44125&-1.4204&0.52185&0.470945\\
          $\lambda_{SH}$&0.60612&0.03331&0.00091576&0.0014208&0.00071561&0.0003719&-0.00022894&-0.00022618&0.00081559&-0.00048164\\
          $\lambda_{DH}$&0.70071&-0.68513&0.0011148&0.0026438&0.00087688&-0.00080729&0.00079511&0.0007723&-0.00078335&0.00096566\\
          $\lambda_{SD}$&-0.018352&0.36301&0.20188&0.23401&-1.8201&1.2962&-2.5734&-2.5342&2.5104&2.4429\\
          $M_{S}\text{/GeV}$&570.9&506.36&69.768&71.562&1240.6&1065.4&1138&1105.3&1285.9&1273.8\\
          $M_{D}\text{/GeV}$&4246.1&1828.5&1385.2&1214.9&1245.6&1096.6&1723.7&1655.3&3126.3&3159.6\\
          $M_{L^{\prime}}\text{/GeV}$&825.16&1529.5&961.08&966.61&1627&2743.7&1155.1&1116.3&1682.8&1654.6\\
          $M_{Q^{\prime}}\text{/GeV}$&1726&4630.8&1987&1922.7&5986.8&3741.7&4368.4&4432.1&1295.4&1295.1\\
          $M_{S_{1}}\text{/GeV}$&586.66&507.25&63.004&63.145&1237.1&1058.8&1138&1105.2&1285.8&1273.8\\
          $M_{S_{2}}\text{/GeV}$&4248.6&1822.8&1385.6&1215.4&1249.1&1102.9&1723.7&1655.3&3126.3&3159.6\\
    \hline
    \end{tabular}
    \captionsetup{justification=raggedright}
    \caption{Benchmark points in DM annihilation and co-annihilation.}
    \label{Benchmark points}
\end{table}

\begin{figure}[h]
	\centering
	\begin{subfigure}{0.45\textwidth}
	   \includegraphics[width=\textwidth]{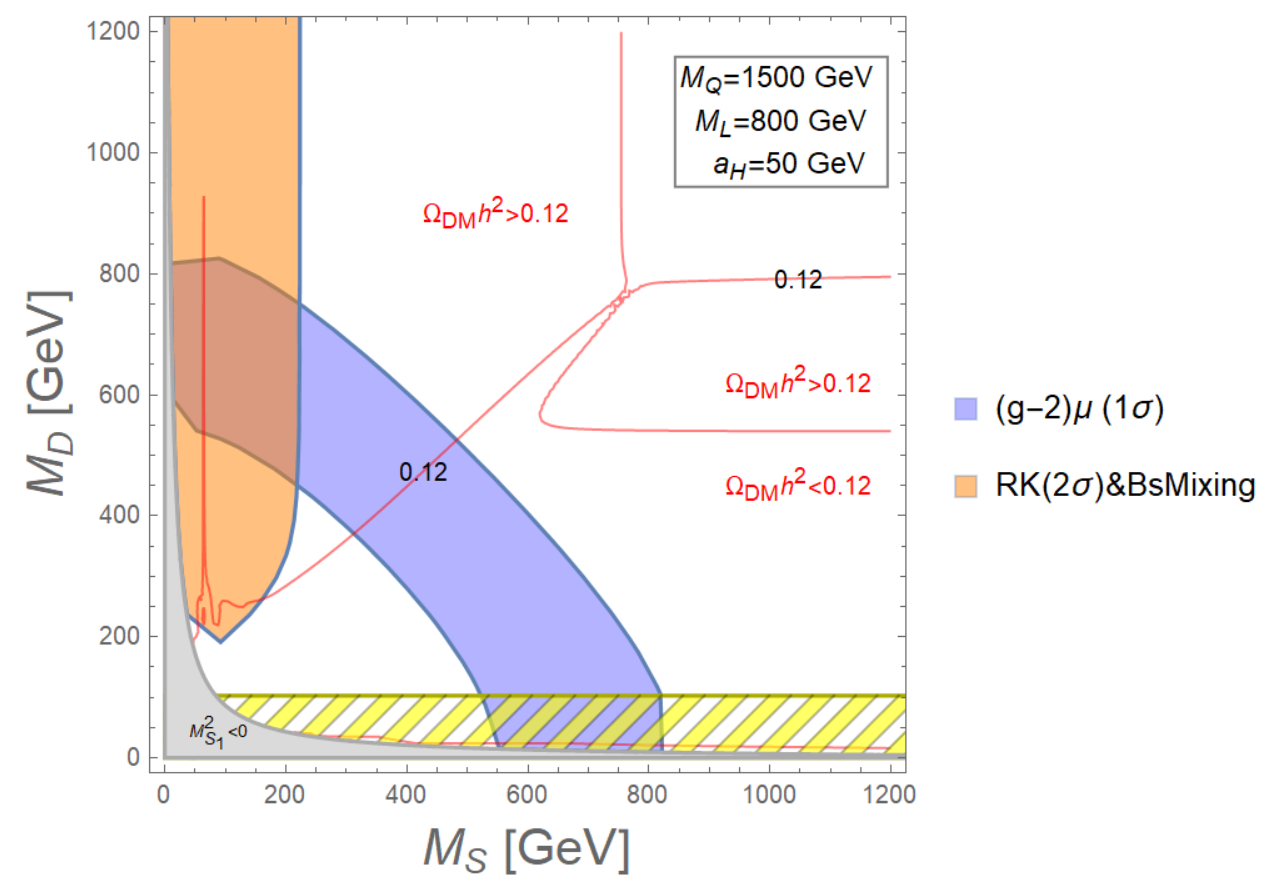}
	\hspace{3em} 
	\end{subfigure}
	\hspace{3em}
	\begin{subfigure}{0.45\textwidth}
	    \includegraphics[width=\textwidth]{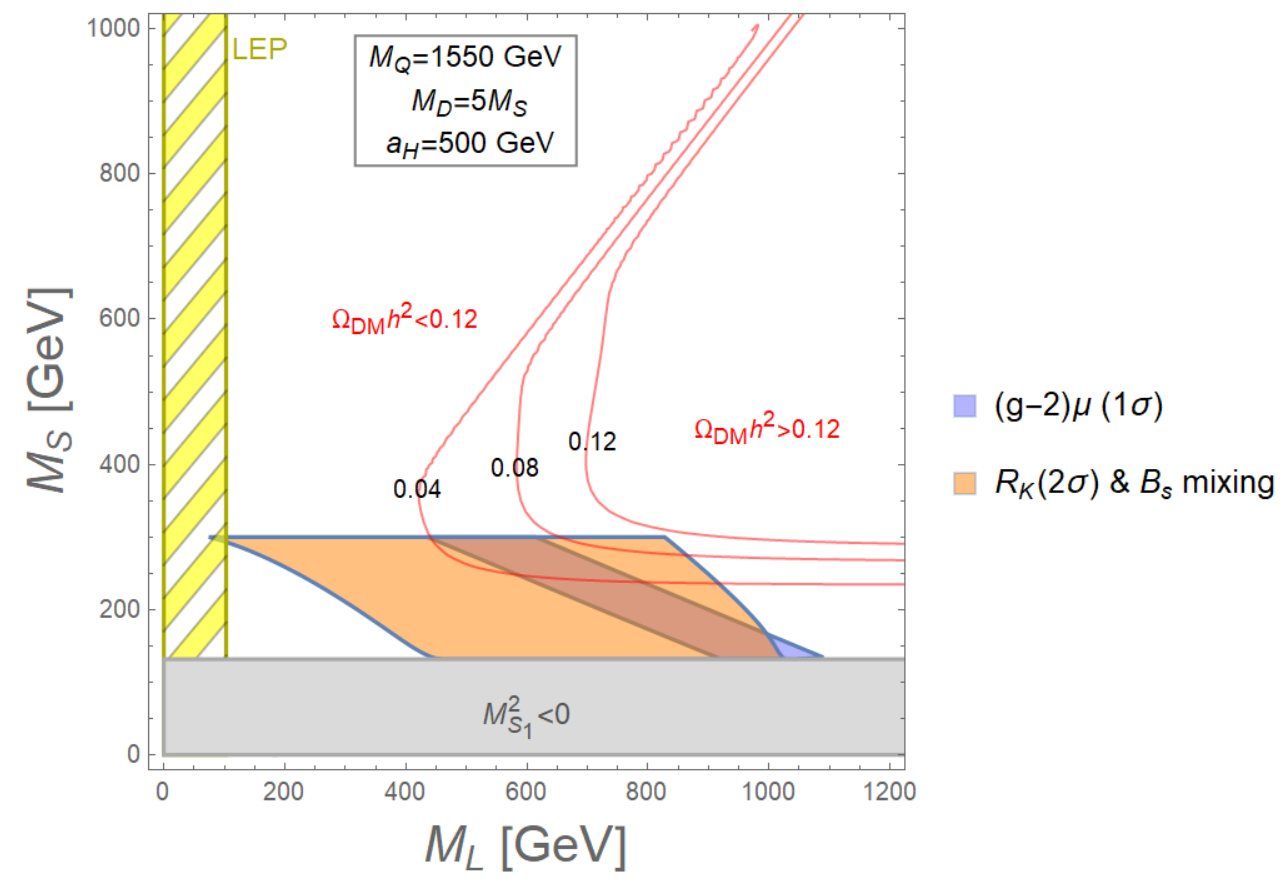}
	\hspace{3em} 
	\end{subfigure}
	\captionsetup{justification=raggedright}
	\caption{The combined constraints on the $M_{S}$ versus $M_{D}$ plane and $M_{L}$ versus $M_{S}$ plane. The couplings are
	$\lambda_{2}^{Q}=0.49$, $\lambda_{3}^{Q}=-0.49$, $\lambda_{2}^{L}=1.4$, $\lambda_{2}^{E}=-0.4$, $\lambda_{SH}=\lambda_{DH}=0.001$ (left),   $\lambda_{2}^{Q}=0.49$, $\lambda_{3}^{Q}=-0.49$, $\lambda_{2}^{L}=1.5$, $\lambda_{2}^{E}=-0.06$, $\lambda_{SH}=\lambda_{DH}=0.001$ (right).}
\label{MSML}
\end{figure}

\begin{figure}[ht]
	\centering
	\begin{subfigure}{0.45\textwidth}
	   \includegraphics[width=\textwidth]{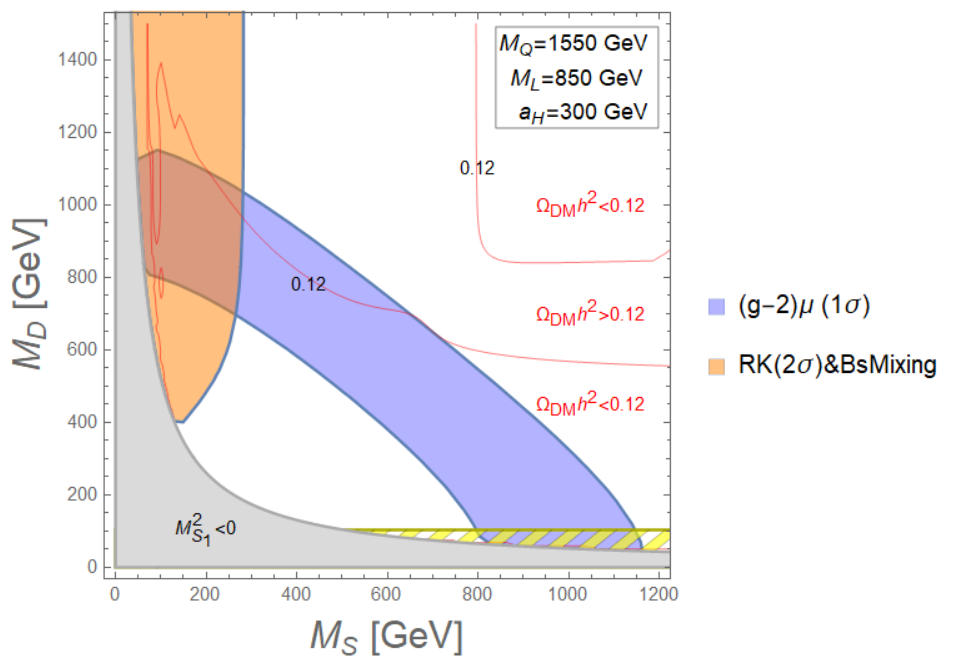}
	\hspace{3em} 
	\end{subfigure}
	\hspace{3em}
	\begin{subfigure}{0.45\textwidth}
	    \includegraphics[width=\textwidth]{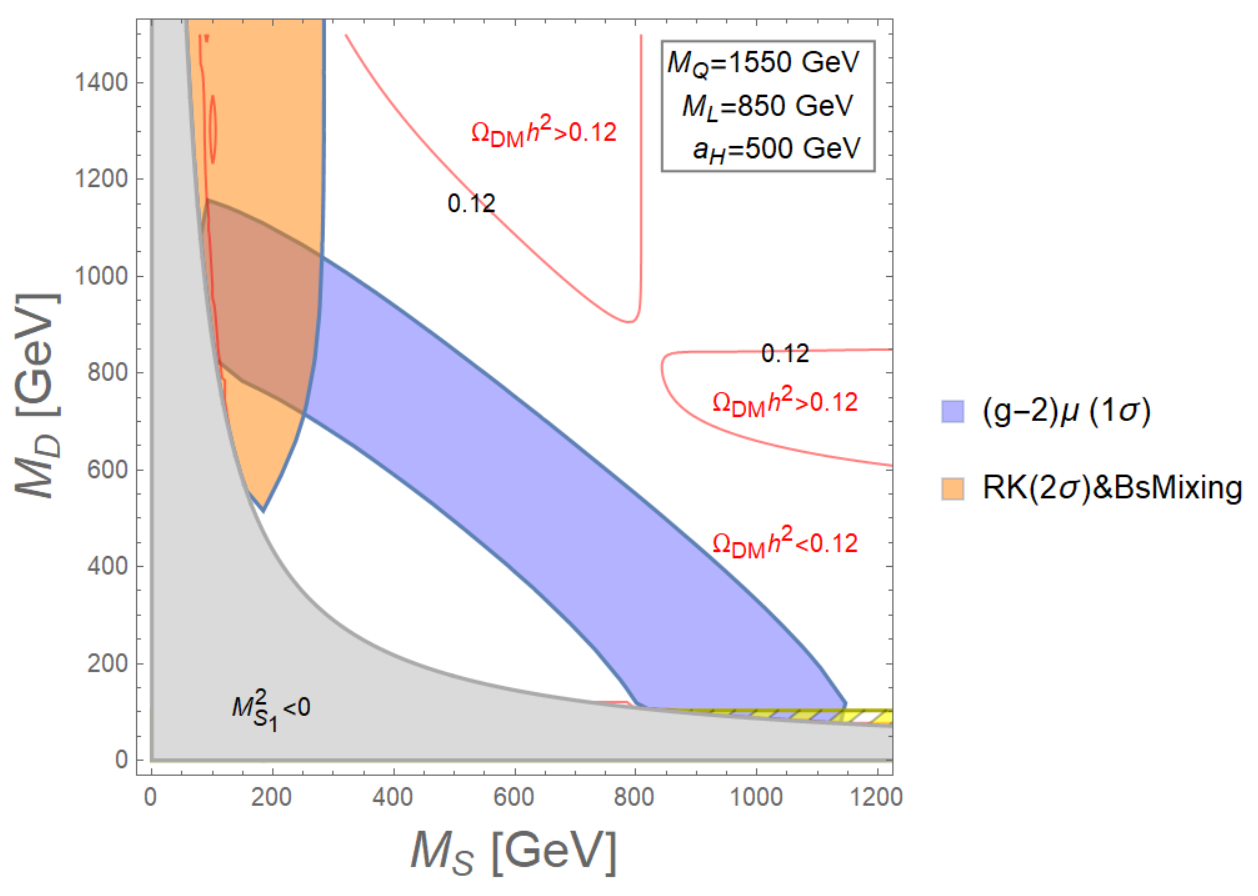}
	\hspace{3em} 
	\end{subfigure}
	\\
	\begin{subfigure}{0.45\textwidth}
	   \includegraphics[width=\textwidth]{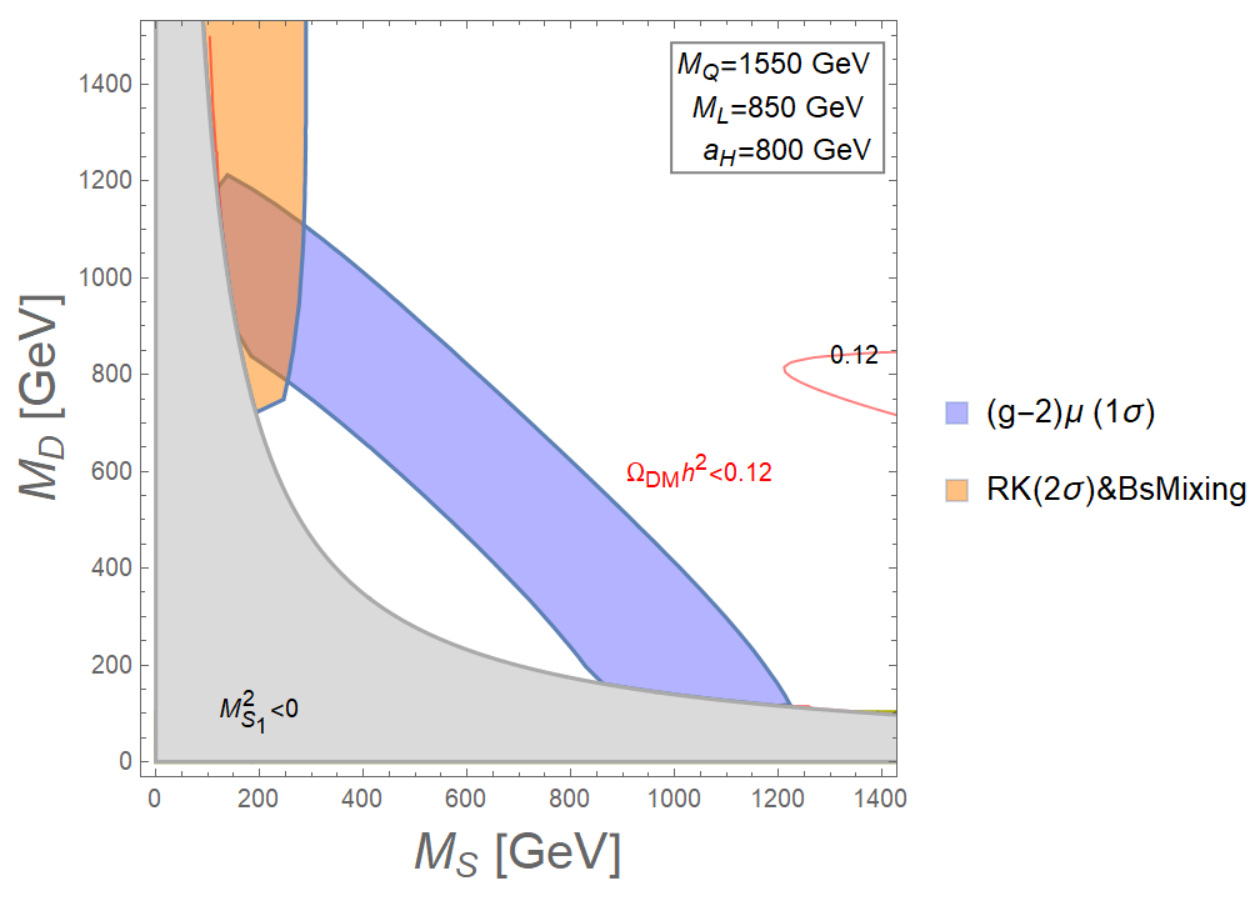}
	\hspace{3em} 
	\end{subfigure}
	\hspace{3em}
	\begin{subfigure}{0.45\textwidth}
	    \includegraphics[width=\textwidth]{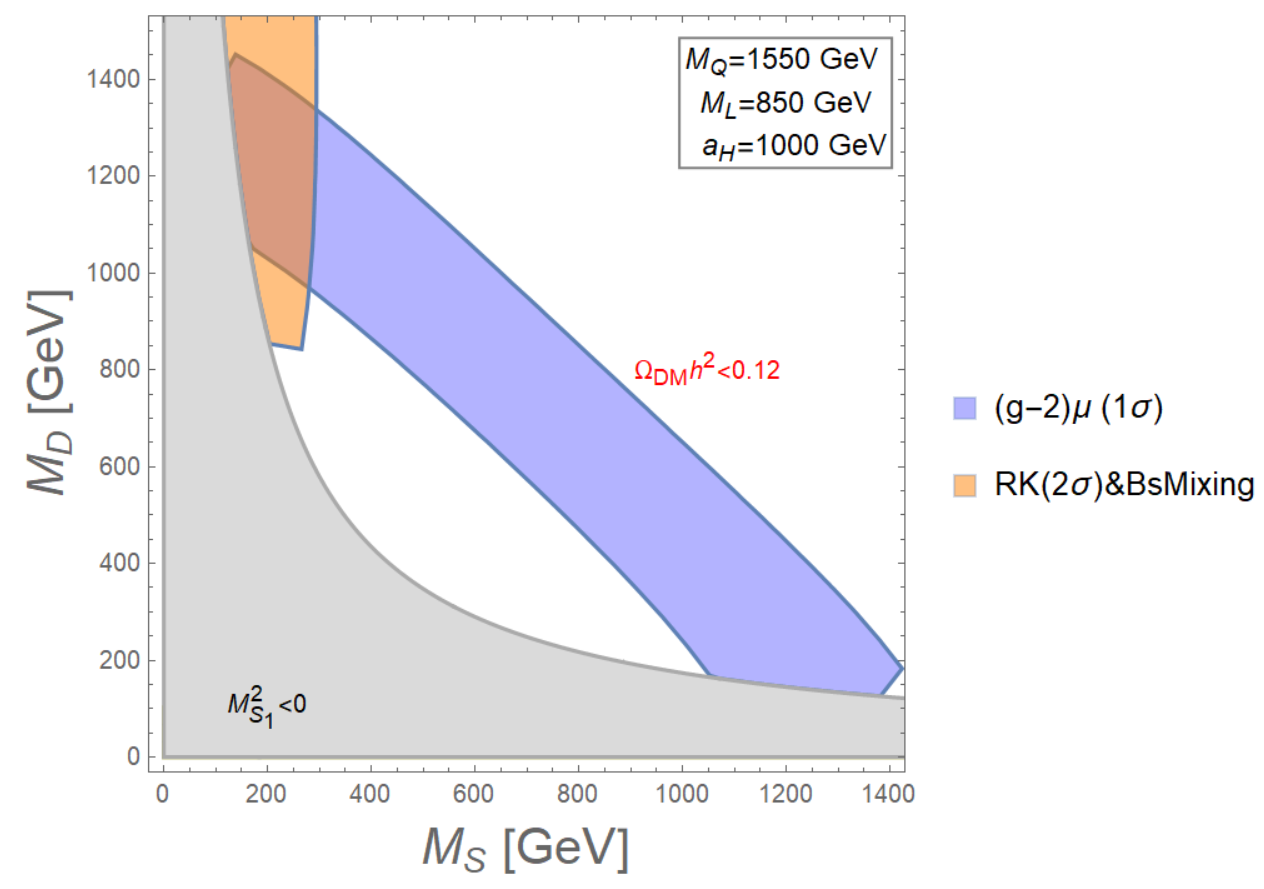}
	\hspace{3em} 
	\end{subfigure}
	\captionsetup{justification=raggedright}
	\caption{The combined constraints on $M_{S}$ and $M_{D}$ plane with different $a_{H}$. The couplings are setting as $\lambda_{2}^{L}=1.5$, $\lambda_{2}^{Q}=0.49$, $\lambda_{3}^{Q}=-0.49$, $\lambda_{SH}=\lambda_{DH}=0.001$ and $\lambda_{2}^{E}$ are -0.1, -0.06, -0.04, -0.04 respectively}
\label{MSMDAH}
\end{figure}

In addition, we investigate the large mixing and relative small coupling case. We can see from Eqs.~\ref{19} and \ref{20}, $\Delta C_{9}$ and $\Delta C_{10}$ are quadratically depended in $\lambda_{2}^{L}$. To reduce the $\lambda_{2}^{L}$, one needs small $M_{Q}$ and $M_{L}$, which are already near the lower edge of experimental search. After a somehow simple analysis, one cannot have a small $\lambda_{2}^{L}$ to account for all the aforementioned anomalies. In the left panel of FIG.~\ref{MSML}, we can see that only Higgs resonance can explain the $R_{K}$, $B_{s}$ mixing, muon $g-2$, and the saturated DM relic density. The right panel of FIG.~\ref{MSML} shows the $S_{1}$ and $L^{\prime}$ co-annihilation process with large mixing ($a_{H}=500$ GeV). In this case, the $R_{K}$, $B_{s}$ mixing, and muon $g-2$ can be explained simultaneously and the DM relic density is undersaturated.
In FIG.~\ref{MSMDAH} we display viable parameter space with different $a_{H}$. In top left panel, the flavor observables and DM relic density can be explained as well. As $a_{H}$ increases, the DM relic density is undersaturated while the constraints of flavour observables are satisfied. 

\begin{figure}[ht]
    \centering
    \includegraphics[width=0.7\textwidth]{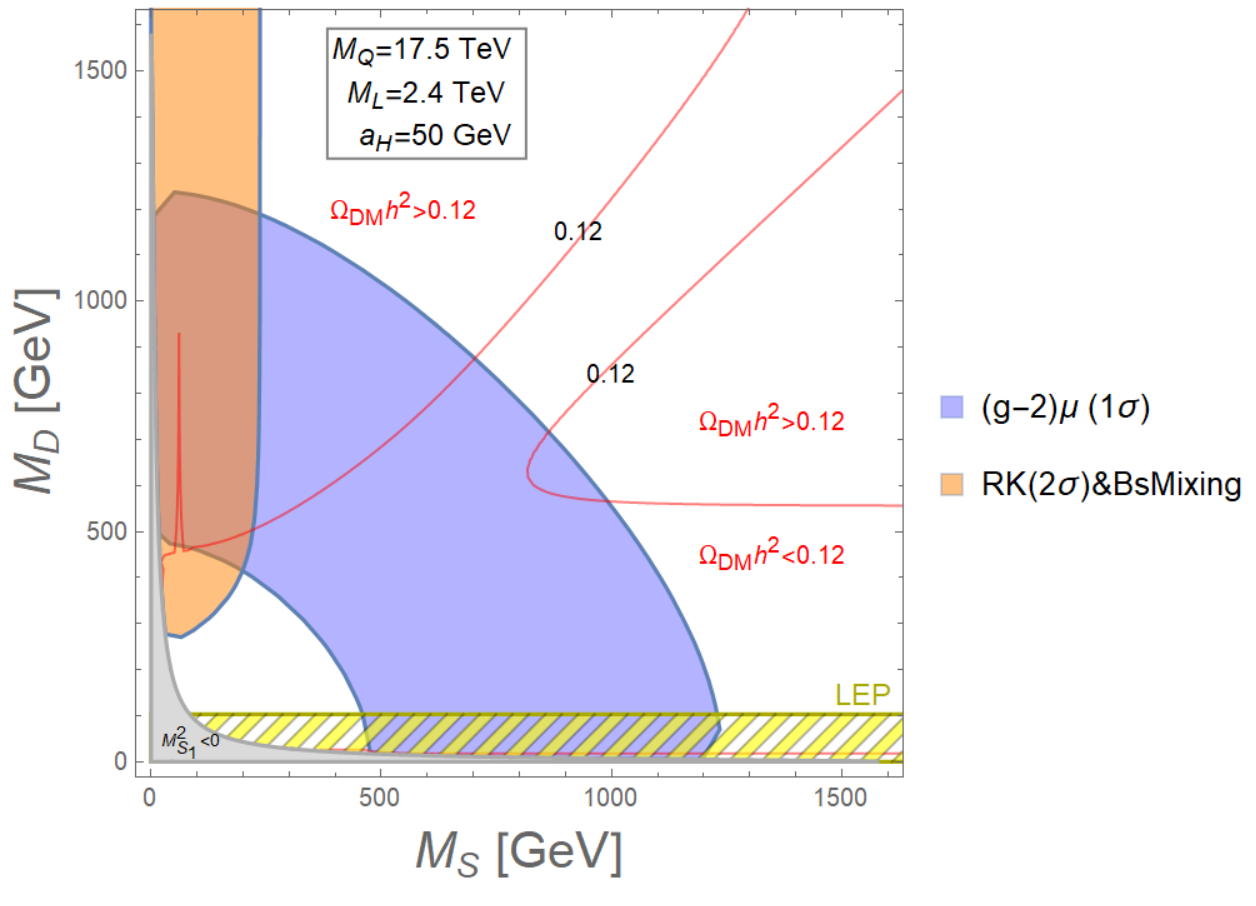}
    \caption{The parameter space consistent with unitarity bounds. The couplings are setting as $\lambda_{2}^{L}=2.9$, $\lambda_{2}^{E}=-2.9$, $\lambda_{2}^{Q}=1.4$, $\lambda_{3}^{Q}=-2.0$, $\lambda_{SH}=\lambda_{DH}=0.001$}
    \label{Unitarity}
\end{figure}

We intend to give some comments on the parameter space on the edge of unitarity. In FIG.~\ref{Unitarity} we employ the conservative parameters which are consistent with the unitarity bounds in Eq.~\ref{16} and avoid large quantum corrections. Roughly speaking, we find the upper bounds on the masess of exotic particles, $M_{Q^{\prime}}=17.5$ TeV and $M_{L^{\prime}}=2.4$ TeV which can account 
for the $R_{K}$, $B_{s}$ mixing, muon $g-2$, and the saturated DM relic density.

\section{$W$ Boson Mass}\label{W mass}
The oblique corrections contributing to $W$ boson mass are given by~\cite{Peskin:1990zt,Peskin:1991sw}
\begin{equation}
 m^{2}_{W}=m^{2}_{W}(SM)+\frac{\alpha c^{2}}{c^{2}-s^{2}}m^{2}_{Z}\left[-\frac{1}{2}\Delta S+c^{2}\Delta T+\frac{c^{2}-s^{2}}{4s^{2}}\Delta U\right]~,   
 \label{30}
\end{equation}
where $\alpha$ is the fine structure constant, c=cos$\theta_{W}$, s=sin$\theta_{W}$. The expressions of $\Delta S$ and $\Delta T$ are
\begin{equation}
\Delta S =\frac{1}{2 \pi}\left[\frac{1}{6} \log \left(\frac{m_{R}^{2}}{m_{S^{\pm}}^{2}}\right)-\frac{5}{36}+\frac{m_{R}^{2} m_{A}^{2}}{3\left(m_{A}^{2}-m_{R}^{2}\right)^{2}} + \frac{m_{A}^{4}\left(m_{A}^{2}-3 m_{R}^{2}\right)}{6\left(m_{A}^{2}-m_{R}^{2}\right)^{3}} \log \left(\frac{m_{A}^{2}}{m_{R}^{2}}\right)\right]~,
\label{31}
\end{equation}

\begin{equation}
\Delta T =\frac{1}{32 \pi^{2} \alpha v^{2}}\left[F\left(m_{S^{\pm}}^{2}, m_{A}^{2}\right)+F\left(m_{S^{\pm}}^{2}, m_{R}^{2}\right)-F\left(m_{A}^{2}, m_{R}^{2}\right)\right]~,
\label{32}
\end{equation}
with the the loop function
\begin{equation}
    F(x, y)= \begin{cases}\frac{x+y}{2}-\frac{x y}{x-y} \log \left(\frac{x}{y}\right), & x \neq y \\ 0, & x=y\end{cases} ~,
\end{equation}
where we ignore $\Delta U$ and rewrite the scalar doublet as
\begin{equation}
    \Phi_{D}=\left(\begin{array}{c}
S_{d}^{0}=R + i A \\
S^{-}
\end{array}\right)
\end{equation}

\begin{figure}[ht]
	\centering
	\begin{subfigure}{0.45\textwidth}
	   \includegraphics[width=\textwidth]{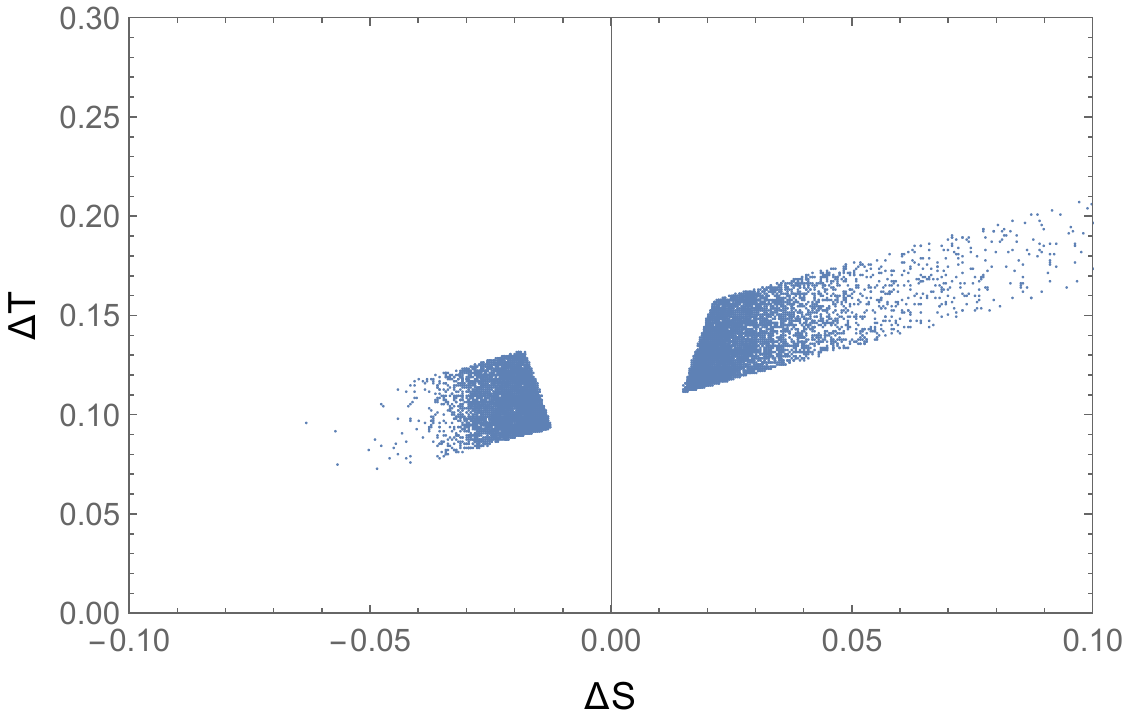}
	\hspace{3em} 
	\end{subfigure}
	\hspace{3em}
	\begin{subfigure}{0.45\textwidth}
	    \includegraphics[width=\textwidth]{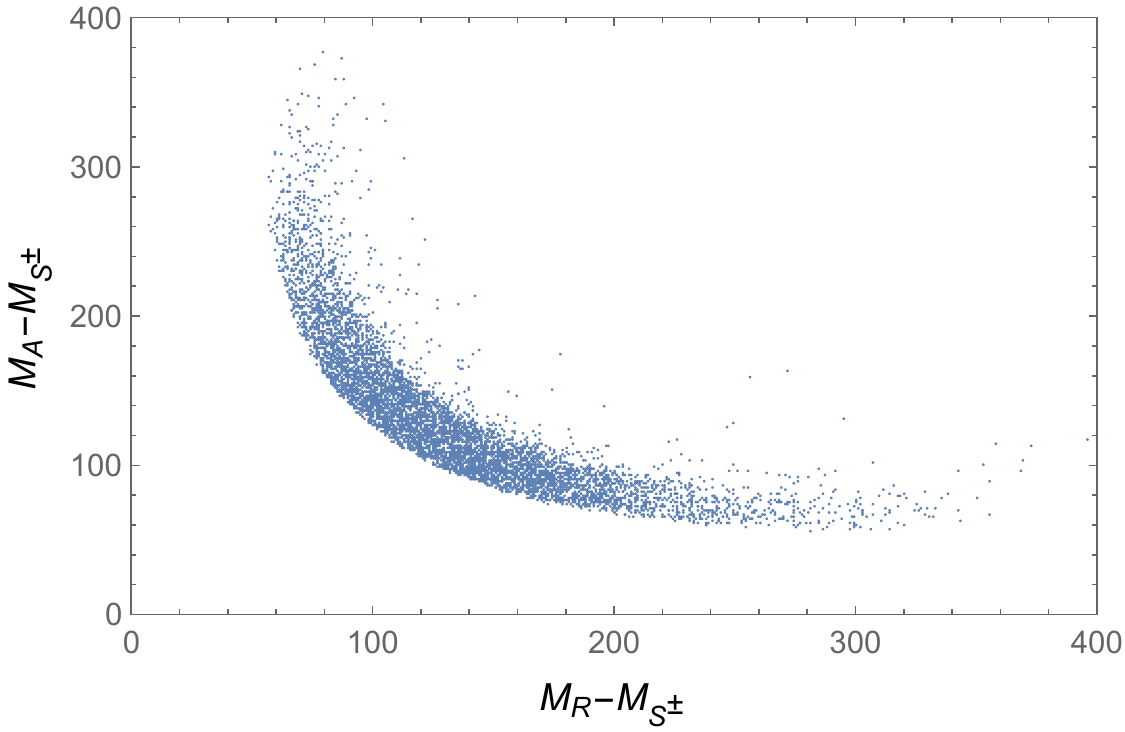}
	\hspace{3em} 
	\end{subfigure}
	\captionsetup{justification=raggedright}
	\caption{The points that explain the up-to-date $W$ boson mass (1 $\sigma$) in $\Delta S$ versus $\Delta T$ plane (left) 
and the mass splitting between neutral and charged parts of the scalar doublet plane (right).}
\label{oblique}
\end{figure}

For convenience, we consider the small mixing between scalar singlet and doublet. To interpret the W boson mass anomaly, the scalar potential is essential to give an appropriate mass splitting between neutral and charged parts or real and imaginary parts of the doublet. The $\lambda_{1}^{\prime}$ term is responsible for the mass splitting between neutral and charged parts, and $\lambda_{2}^{\prime}$ term is responsible for the mass splitting between real and imaginary parts. $\lambda_{DH}$ term provides a shift of $M_{D}$ when Electro-weak symmetry breaking (EWSB).
We find that W boson mass cannot be explained if we consider $\lambda_{2}^{\prime}$ only. However, if the $\lambda_{DH}$, ${\lambda_{2}^{\prime}}$, $\lambda_{1}^{\prime}$ are involved simultaneously, the W mass amomaly can be account for in a straightforward way. We can fit the up-to-date measurement of $W$ boson mass (1 $\sigma$) by using Eqs. \ref{30}, \ref{31} and \ref{32}.
 FIG.\ref{oblique} shows the corresponding oblique parameters between $\Delta S$ and $\Delta T$, and the mass splitting between neutral and charged parts of the scalar doublet.

We would like to point out that  the vectorlike fermion in our model cannot contribute to the oblique parameters in renormalizable level, since the absence of mass splitting between neutral and charged parts.

\section{Discussion and Conclusion}\label{Conclusion}

We have revisited the SESM which can address various tentative new physics anomalies and DM.
 The mass splitting between the charged and neutral parts of scalar doublet via scalar potential can account for the $W$ boson mass anomaly. 
Moreover, we considered the unitarity constraints in the Yukawa sector. Also, we employed the random scan approach,
and obtained the viable parameter spaces which can explain  the B physics anomaly, muon anomalous magnetic moment, 
$W$ mass anomaly, and dark matter relic density simultaneously.
To be concrete,  we select some benchmark points.
The various DM (co)-annihilation and resonance processes are investigated, and the benchmark points whose DM relic density are around
or smaller than the observed DM relic density are demonstrated. In particular, the Higgs pole,
$S_{1}$ and $S_{1}$ annihilation,  $S_{1}$ and  $S_{2}$, $L^{\prime}$, $Q^{\prime}$ co-annihilation are demonstrated as well, 
and all these points can evade the current flavor, XENONnT, and LUX-ZEPLIN  constraints.
To be complete,  we briefly comment on the potential of the SESM. For instance, similar to 
the New Minimal SM (NMSM)~\cite{Davoudiasl:2004be}, 
we can introduce two right-handed neutrinos to explain the neutrino masses and mixing as well as baryon asymmetry, 
and introduce a real scalar to address inflation. Because of the strong constraint on 
the tensor-to-scalar ratio, we need to choose the proper inflation potential, for example,
the inflation potentials in Section 2 of Ref.~\cite{Sabir:2019xwk}.


\begin{acknowledgments}

We would like to thank Lorenzo Calibbi and Jibo He very much for the helpful discussions. 
This work is supported in part by 
the National Key Research and Development Program of China Grant No. 2020YFC2201504, 
by the Projects No. 11875062, No. 11947302, and No. 12047503 supported 
by the National Natural Science Foundation of China,
as well as by the Key Research Program 
of the Chinese Academy of Sciences, Grant No. XDPB15.

\end{acknowledgments}

\appendix
\section{The Unitarity Constraint on $\lambda_{2}^{Q}$}{\label{A}}
\subsection{The Unitarity Constraint on $\lambda_{2}^{Q}$}
\subsubsection{$J=0$ Partial Wave}

Two particle system is defined as
\begin{equation}
    \begin{aligned}
        &|\psi \bar{\psi}\rangle_{1}=\frac{1}{\sqrt{N}} \delta_{a}^{b}\left|\psi_{a} \bar{\psi}^{b}\right\rangle ~,\\
        &|\psi \bar{\psi}\rangle_{\text {Adj }}^{A}=\sqrt{2}\left(T^{A}\right)_{a}^{\cdot b}\left|\psi_{a} \bar{\psi}^{b}\right\rangle ~,\\
        &|\psi \psi\rangle_{\mathbf{S}}^{A}=\left(T_{\mathrm{S}}^{A}\right)_{a b}\left|\psi_{a} \psi_{b}\right\rangle~.
        \end{aligned}
    \end{equation}

The group factor can be calculated as follows
\begin{equation}
    \begin{split}
    \langle \psi \bar{\psi}|\psi \bar{\psi}\rangle_{1}&=\langle \psi_{c} \bar{\psi^{d}}| \frac{1}{\sqrt{N}} \delta_{c}^{d}\frac{1}{\sqrt{N}} \delta_{a}^{b}\left|\psi_{a} \bar{\psi}^{b}\right\rangle\\
    &=\frac{N}{\sqrt{N}} \frac{N}{\sqrt{N}}\langle \psi_{1} \bar{\psi}_{1}|\psi_{1} \bar{\psi}_{1}\rangle\\
    &=N\langle \psi_{1} \bar{\psi}_{1}|\psi_{1} \bar{\psi}_{1}\rangle~,
    \end{split}
\end{equation}
where $\langle \psi \bar{\psi}|\psi \bar{\psi}\rangle_{1}$ denotes the elastic scattering process in singlet channel, 
and $\psi_{1}$ stands for one component of $\psi$ in the fundamental representation.

\subsubsection{$J=\frac{1}{2}$ Partial Wave}

$\mathcal{T}^{+0^{\star}+0^{\star}}$ and $\mathcal{T}^{-0-0}$ contribute to $J=\frac{1}{2}$ partial wave as follows
\begin{equation}
    +0^{*}\left\{\begin{array}{l}
        \overline{Q}_{2} \Phi_{S}^{\star} \sim \overline{\square}\\
        \end{array}\right.~.
\end{equation}
The one fermion and one scalar state is
\begin{equation}
    |\Psi S \rangle = 1 |\Psi S \rangle~,
\end{equation}
with group factor
\begin{equation}
    \begin{split}
       \langle \Psi S |\Psi S \rangle=1\langle \Psi S |\Psi S \rangle~.
    \end{split}
\end{equation}

We have $+0^{\star}+0^{\star} \text{or} -0-0\left\{\mathcal{F}_{\overline{Q}_{2} \Phi^{\star}_{S} \overline{Q}_{2} \Phi^{\star}_{S}}^{u,\overline{\square}}=\mathcal{F}_{Q_{2} \overline{\Phi}^{\star}_{S} Q_{2} \overline{\Phi}^{\star}_{S}}^{u,\square}=1\right.$. In $(\overline{Q}_{2} \Phi^{\star}_{S},Q_{2} \overline{\Phi}^{\star}_{s})$ basis, we have (Note that the base $(Q_{R}^{\prime} \Phi_{s}^{\star},\overline{Q}_{R}^{\prime} \Phi_{s})$ leads to the same result.)
\begin{equation}
    \begin{split}
    a_{S U(3)=\bar{\square} or \square, S U(2)=\bar{\square} or \square}^{J=1/2}&=\frac{{\lambda_{2}^{Q}}^{2}}{32 \pi} \int_{-1}^{+1} \mathrm{~d} \cos \theta d_{\frac{1}{2} \frac{1}{2}}^{\frac{1}{2}}(\theta)\left(\begin{array}{cc}
         \mathcal{T}_{u}^{+0^{\star}+0^{\star}} & 0 \\
        0 &  \mathcal{T}_{u}^{-0-0}
        \end{array}\right)\\
        &=\frac{{\lambda_{2}^{Q}}^{2}}{32 \pi} \int_{-1}^{+1} \mathrm{~d} \cos \theta \cos \frac{\theta}{2}\left(\begin{array}{cc}
            -\frac{1}{\cos \frac{\theta}{2}} & 0 \\
           0 &  -\frac{1}{\cos \frac{\theta}{2}}
           \end{array}\right)\\
        &=-\frac{{\lambda_{2}^{Q}}^{2}}{16 \pi}~.
    \end{split}
\end{equation}

This corresponds to the bound
\begin{equation}
    |\lambda_{2}^{Q}| \leq 5.01~.
\end{equation}

\subsection{The Unitarity Constraint on $\lambda_{3}^{Q}$}

The $\lambda_{3}^{Q}$ in Eq.~\ref{lagrangian} is the first model of dirac type theory in both $SU(3)_C$ and $SU(2)_L$ as well, and thus the constraint on $\lambda_{3}^{Q}$ is exact the same as $\lambda_{2}^{Q}$. We have the strict bound $|\lambda_{3}^{Q}| \leq 2.05$ from $J=0$ partial wave. 

\subsection{The Unitarity Constraint on $\lambda_{2}^{L}$}

As we have claimed in the quark sector before, the Lagrangian becomes $\lambda_{i}^{L} \overline{L^{\prime}} L_{i} \Phi_{S} \rightarrow \lambda_{2}^{L} \overline{L^{\prime}}_{R} L_{2} \Phi_{S}$. This kind of interaction can be classified by the first type of dirac theory 
in $SU(2)_L$.

\subsubsection{$J=0$ Partial Wave}

One can consider the elastic scatting processes $L_{R}^{\prime} \overline{L}_{2} \rightarrow L_{R}^{\prime} \overline{L}_{2}$ or $\overline{L}_{R}^{\prime} {L}_{2L} \rightarrow \overline{L}_{R}^{\prime} {L}_{2}$. Group factor is  $++++ or ----\left\{\mathcal{F}_{L_{R}^{\prime} \overline{L}_{2} L_{R}^{\prime} \overline{L}_{2}}^{s,1}=\mathcal{F}_{\overline{L}_{R}^{\prime} {L}_{2} \overline{L}_{R}^{\prime} {L}_{2}}^{s,1}=N\right.$.

In $(L^{\prime}_{R} \overline{L}_{2},\overline{L}^{\prime}_{R} L_{2})$ basis, we have
\begin{equation}
    \begin{split}
    a_{ S U(2)=1}^{J=0}&=\frac{{\lambda_{2}^{L}}^{2}}{32 \pi} \int_{-1}^{+1} \mathrm{~d} \cos \theta d_{00}^{0}(\theta)\left(\begin{array}{cc}
        N_{2} \mathcal{T}_{s}^{++++} & 0 \\
        0 & N_{2} \mathcal{T}_{s}^{----}
        \end{array}\right)\\
        &=\frac{{\lambda_{2}^{L}}^{2}}{32 \pi} \int_{-1}^{+1} \mathrm{~d} \cos \theta \left(\begin{array}{cc}
            2\times (-1) & 0 \\
            0 & 2\times(-1)
            \end{array}\right)\\
        &=-\frac{{\lambda_{2}^{L}}^{2}}{8 \pi}~,
    \end{split}
\end{equation}
which gives the bound
\begin{equation}
    |\lambda_{2}^{L}|<3.54~.
\end{equation}

\subsubsection{$J=\frac{1}{2}$ Partial Wave}

Consider the elastic scatting process $L^{\prime}_{R} \Phi^{\star}_{S} \rightarrow L^{\prime}_{R} \Phi^{\star}_{S}$ or $\overline{L}^{\prime}_{R} \Phi_{S} \rightarrow \overline{L}^{\prime}_{R} \Phi_{S}$, group factor is  $+0^{\star}+0^{\star} or - 0 - 0\left\{\mathcal{F}_{L^{\prime}_{R} \Phi^{\star}_{S} L^{\prime}_{R} \Phi^{\star}_{S}}^{s,\square}=\mathcal{F}_{\overline{L}^{\prime}_{R} \Phi_{S} \overline{L}^{\prime}_{R} \Phi_{S}}^{s,\overline{\square}}=1\right.$.
In $(L^{\prime}_{R} \Phi^{\star}_{S},\overline{L}^{\prime}_{R} \Phi_{S})$ basis, we get (the basis $(\overline{L}_{2} \Phi_{S}^{\star},L_{2} \Phi_{S})$ gives the same result.)
\begin{equation}
    \begin{split}
    a_{ S U(2)=\square or \overline{\square}}^{J=1/2}&=\frac{{\lambda_{2}^{L}}^{2}}{32 \pi} \int_{-1}^{+1} \mathrm{~d} \cos \theta d_{\frac{1}{2} \frac{1}{2}}^{\frac{1}{2}}(\theta)\left(\begin{array}{cc}
        \mathcal{T}_{u}^{+0^{\star}+0^{\star}} & 0 \\
        0 & \mathcal{T}_{u}^{-0-0}
        \end{array}\right)\\
        &=\frac{{\lambda_{2}^{L}}^{2}}{32 \pi} \int_{-1}^{+1} \mathrm{~d} \cos \theta \cos \frac{\theta}{2} \left(\begin{array}{cc}
            -\frac{1}{\cos \frac{\theta}{2}} & 0 \\
            0 & -\frac{1}{\cos \frac{\theta}{2}}
            \end{array}\right)\\
        &=-\frac{{\lambda_{2}^{L}}^{2}}{16 \pi}~,
    \end{split}
\end{equation}
with the relatively weak bound
\begin{equation}
    |\lambda_{2}^{L}| < 5.01~.
\end{equation}

\subsection{The Unitarity Constraint on $\lambda_{2}^{E}$}

The Lagrangian is given by $\lambda_{i}^{E} \overline{L}^{\prime} E_{i} \widetilde{\Phi}_{D} \rightarrow \lambda_{2}^{E} \overline{L}^{\prime}_{L} E_{2} \widetilde{\Phi}_{D}$. Since $E$ is the $SU(2)_L$ singlet and $\Phi_{D}$ is $SU(2)_L$ doublet, this kind of interaction can be classified into the second type of dirac theory.
\subsubsection{$J=0$ Partial Wave}
Two particle state is defined (E is $SU(2)_L$ singlet fermion) as
\begin{equation}
    |\bar{L}_{L}^{\prime} E\rangle_{\overline{\square}}^{a}=\left|\bar{L}_{L}^{\prime a} E\right\rangle~.
\end{equation}
The group factor is
\begin{equation}
    \langle\bar{L}_{L}^{\prime a} E \left|\bar{L}_{L}^{\prime a} E\right\rangle=1\langle\bar{L}_{L}^{\prime a} E \left|\bar{L}_{L}^{\prime a} E\right\rangle~.
\end{equation}
We have $++++ or ----\left\{\mathcal{F}_{\overline{L}^{\prime}_{L} E \overline{L}^{\prime}_{L} E}^{s,\overline{\square}}=\mathcal{F}_{L^{\prime}_{L} \overline{E} L^{\prime}_{L} \overline{E}}^{s,\square}=1\right.$.

In $(\overline{L}^{\prime}_{L} E, L^{\prime}_{L} \overline{E})$ basis, we have
\begin{equation}
    \begin{split}
    a_{ S U(2)=\bar{\square} or \square}^{J=0}&=\frac{{\lambda_{2}^{E}}^{2}}{32 \pi} \int_{-1}^{+1} \mathrm{~d} \cos \theta d_{00}^{0}(\theta)\left(\begin{array}{cc}
         \mathcal{T}_{s}^{++++} & 0 \\
        0 &  \mathcal{T}_{s}^{----}
        \end{array}\right)\\
        &=\frac{{\lambda_{2}^{E}}^{2}}{32 \pi} \int_{-1}^{+1} \mathrm{~d} \cos \theta \left(\begin{array}{cc}
            -1 & 0 \\
           0 &  -1
           \end{array}\right)\\
        &=-\frac{{\lambda_{2}^{E}}^{2}}{16 \pi}~,
    \end{split}
\end{equation}
with unitarity bound
\begin{equation}
    |\lambda_{2}^{E}| < 5.01~.
\end{equation}

\subsubsection{$J=\frac{1}{2}$ Partial Wave}

Group factor is  $+0+0$ or $-0^{\star}-0^{\star}\left\{\mathcal{F}_{E \widetilde{\Phi}_{D} E \widetilde{\Phi}_{D}}^{s,\square}=\mathcal{F}_{\overline{E} \widetilde{\Phi}^{\star}_{D} \overline{E} \widetilde{\Phi}^{\star}_{D}}^{s,\overline{\square}}=1\right.$.
In $(E \widetilde{\Phi}_{D},\overline{E} \widetilde{\Phi}^{\star}_{D})$ base, we
have (We should claim that $\widetilde{\Phi}_{D}$ is in the fundamental representation.)
\begin{equation}
    \begin{split}
    a_{ S U(2)=\square or \overline{\square}}^{J=1/2}&=\frac{{\lambda_{2}^{E}}^{2}}{32 \pi} \int_{-1}^{+1} \mathrm{~d} \cos \theta d_{\frac{1}{2} \frac{1}{2}}^{\frac{1}{2}}(\theta)\left(\begin{array}{cc}
         \mathcal{T}_{s}^{+0+0} & 0 \\
        0 &  \mathcal{T}_{s}^{-0^{\star}-0^{\star}}
        \end{array}\right)\\
        &=\frac{{\lambda_{2}^{E}}^{2}}{32 \pi} \int_{-1}^{+1} \mathrm{~d} \cos \theta \cos \frac{\theta}{2}\left(\begin{array}{cc}
            -\cos \frac{\theta}{2} & 0 \\
           0 &  -\cos \frac{\theta}{2}
           \end{array}\right)\\
        &=-\frac{{\lambda_{2}^{E}}^{2}}{32 \pi}~,
    \end{split}
\end{equation}
with unitarity bound 
\begin{equation}
    |\lambda_{2}^{E}|< 7.09~.
\end{equation}
However, an alternative basis $(\overline{L}^{\prime}_{L} \widetilde{\Phi}_{D},L^{\prime}_{L} \widetilde{\Phi}_{D}^{\star})$ gives 
stronger constraint since the group factor is $+0+0$ or $-0^{\star}-0^{\star}\left\{\mathcal{F}_{\overline{L}^{\prime}_{L} \widetilde{\Phi}_{D} \overline{L}^{\prime}_{L} \widetilde{\Phi}_{D}}^{s,1}=\mathcal{F}_{L^{\prime}_{L} \widetilde{\Phi}_{D}^{\star} L^{\prime}_{L} \widetilde{\Phi}_{D}^{\star}}^{s,1}=N\right.$. In this case, we get
\begin{equation}
    \begin{split}
    a_{ S U(2)=1}^{J=1/2}&=\frac{{\lambda_{2}^{E}}^{2}}{32 \pi} \int_{-1}^{+1} \mathrm{~d} \cos \theta d_{\frac{1}{2} \frac{1}{2}}^{\frac{1}{2}}(\theta)\left(\begin{array}{cc}
         N_{2} \mathcal{T}_{s}^{+0+0} & 0 \\
        0 & N_{2} \mathcal{T}_{s}^{-0^{\star}-0^{\star}}
        \end{array}\right)\\
        &=\frac{{\lambda_{2}^{E}}^{2}}{32 \pi} \int_{-1}^{+1} \mathrm{~d} \cos \theta \cos \frac{\theta}{2}\left(\begin{array}{cc}
            2\times(-\cos \frac{\theta}{2}) & 0 \\
           0 &  2\times(-\cos \frac{\theta}{2})
           \end{array}\right)\\
        &=-\frac{{\lambda_{2}^{E}}^{2}}{16 \pi}~,
    \end{split}
\end{equation}
with the group factor enhancement, the unitarity bound becomes
\begin{equation}
    |\lambda_{2}^{E}|<5.01~.
\end{equation}


\end{document}